\documentclass[aps,prc,preprint,superscriptaddress,groupedaddress,
showpacs,floatfix]{revtex4}
\usepackage{graphicx}

\begin{document}

\title{Charge and matter distributions and form factors of light, medium and
heavy neutron-rich nuclei}

\author{A.~N.~Antonov}
\affiliation{Institute for Nuclear Research and Nuclear Energy,
Bulgarian Academy of Sciences, Sofia 1784, Bulgaria}

\author{D.~N.~Kadrev}
\affiliation{Institute for Nuclear Research and Nuclear Energy,
Bulgarian Academy of Sciences, Sofia 1784, Bulgaria}

\author{M.~K.~Gaidarov}
\affiliation{Institute for Nuclear Research and Nuclear Energy,
Bulgarian Academy of Sciences, Sofia 1784, Bulgaria}

\author{E.~Moya de Guerra}
\affiliation{Instituto de Estructura de la Materia, CSIC, Serrano
123, 28006 Madrid, Spain}

\author{P.~Sarriguren}
\affiliation{Instituto de Estructura de la Materia, CSIC, Serrano
123, 28006 Madrid, Spain}

\author{J.~M.~Udias}
\affiliation{Departamento de Fisica Atomica, Molecular y Nuclear,
Facultad de Ciencias Fisicas, Universidad Complutense de
Madrid, Madrid E-28040, Spain}

\author{V.~K.~Lukyanov}
\affiliation{Joint Institute for Nuclear Research, Dubna 141980,
Russia}

\author{E.~V.~Zemlyanaya}
\affiliation{Joint Institute for Nuclear Research, Dubna 141980,
Russia}

\author{G.~Z.~Krumova}
\affiliation{University of Rousse, Rousse 7017, Bulgaria}


\begin{abstract}
Results of charge form factors calculations for several unstable
neutron-rich isotopes of light, medium and heavy nuclei (He, Li,
Ni, Kr, Sn) are presented and compared to those of stable isotopes
in the same isotopic chain. For the lighter isotopes (He and Li)
the proton and neutron densities are obtained within a microscopic
large-scale shell-model, while for heavier ones Ni, Kr and Sn the
densities are calculated in deformed self-consistent mean-field
Skyrme HF+BCS method. We also compare proton densities to matter
densities together with their rms radii and diffuseness parameter
values. Whenever possible comparison of form factors, densities
and rms radii with available experimental data is also performed.
Calculations of form factors are carried out both in plane wave
Born approximation (PWBA) and in distorted wave Born approximation
(DWBA). These form factors are suggested as predictions for the
future experiments on the electron-radioactive beam colliders
where the effect of the neutron halo or skin on the proton
distributions in exotic nuclei is planned to be studied and
thereby the various theoretical models of exotic nuclei will be
tested.
\end{abstract}
\maketitle

\section{Introduction}

The scattering of particles and ions from nuclei has provided
along the years invaluable information on charge, matter, current
and momentum distributions of stable isotopes. At present, efforts
are devoted to investigate with such probes highly unstable
isotopes at radioactive nuclear beam (RNB) facilities. Since the
first experiments \cite{Tan85a,Tan85b,Tan92,Tan85c,Tan88a,Tan88b}
it has been found from analyses of total interaction cross
sections that weakly-bound neutron-rich light nuclei, e.g.
$^{6,8}$He, $^{11}$Li, $^{14}$Be, $^{17,19}$B, have increased
sizes that deviate substantially from the $R\sim A^{1/3}$ rule. It
was realized (e.g. \cite{Han95,Dob94,Cas00}) that such a new
phenomenon is due to the weak binding of the last few nucleons
which form a diffuse nuclear cloud due to quantum-mechanical
penetration (the so called "nuclear halo"). Another effect is that
the nucleons can form a "neutron skin" \cite{Tani95} when the
neutrons are on average less bound than the protons. The origin of
the skin lies in the large difference of the Fermi energy levels
of protons and neutrons so that the neutron wave function extends
beyond the effectively more bound proton wave function
\cite{Cas00}. Thus, the term "neutron skin" describes an excess of
neutrons at the nuclear surface, whereas the "halo" stands for
such excess plus a long tail of the neutron density distribution.

Most exotic nuclei are so shortlived that they cannot be used as
targets at rest. Instead, direct reactions with RNB can be done in
inverse kinematics, where the role of beam and target are
interchanged. For example, proton elastic scattering angular
distributions were measured at incident energies less than 100
MeV/nucleon for He isotopes (e.g.
\cite{Oga99a,Oga99b,Cor97,Lag01,Cor96,Kor97a,Kor97b,Ter01,Chu95,
Kor93}) and Li isotopes (e.g. \cite{Cas00,Kor97a}) and at an
energy of 700 MeV/nucleon for the same nuclei at GSI (Darmstadt)
(e.g. \cite{Alk02,Ege01,Ege02,Neu02,Ege03}). The charge and matter
distributions of these nuclei were tested in analyses of
differential and total reaction cross sections of the proton
scattering on exotic nuclei using different phenomenological and
theoretical methods (see
\cite{Kor97a,Kor97b,Chu95,Cre95,Zhu93,Zhu94,Avr00,Avr02,
Dor98,Alk02,Ege01,Ege02,Ege03,Neu02,Kar97}). It was shown (e.g.
\cite{Avr02}) that elastic scattering of protons serves as a good
tool to distinguish between different models of density
distributions. It was demonstrated for the case of intermediate
incident energies that proton scattering in the region of small
momentum transfer is particularly sensitive to the nuclear matter
radius and the halo structure of nuclei \cite{Ege03}.

The elastic proton scattering experiments for studying the
$^{6,8}$He, $^{8,9,11}$Li isotopes have been performed at GSI by
using external targets. As noted in \cite{Ege03}, however, the use
of internal targets at storage rings in the new generation
radioactive beam facilities will have advantage over external
target experiments and will allow to extend such investigations to
a wide range of medium and heavy nuclei.

Concerning the charge distributions of nuclei, it is known that
their most accurate determination can be obtained from
electron-nucleus scattering. For the case of exotic nuclei the
corresponding charge densities are planned to be obtained by
colliding electrons with these nuclei in storage rings. As shown
in the NuPECC Report \cite{NuP00}, a first technical proposal for
a low-energy electron-heavy-ion collider made at JINR (Dubna) has
been further developed and incorporated in the GSI physics program
\cite{Shr01} along with the plan for the electron-ion collider at
the MUSES facility at RIKEN \cite{Sud01,Kat03}. Several
interesting and challenging issues can be analyzed by the
mentioned electron scattering experiments. One of them is to study
how the charge distribution evolves with increasing neutron number
(or isospin) at fixed proton number. The question remains up to
what extent the neutron halo or skin may trigger sizable changes
of the charge root-mean-square (rms) radius, as well as of the
diffuseness in the peripherical region of the charge distribution.
This point may then be very important for understanding the
neutron-proton interaction in the nuclear medium. To this end the
preliminary theoretical calculations of the charge form factors of
neutron-rich exotic nuclei can serve as a challenge for future
experimental works and thus, for accurate determination of the
charge distributions in these nuclei. This can be a test of the
different theoretical models used for predicting charge
distributions.

In recent years theoretical work has been done along these lines
focusing on halo nuclei (e.g. \cite{Gar99,Gar00,Moy02,Ant04}). In
\cite{Gar00,Moy02} the Borromean nuclei are described as
three-body systems and the electron-ion scattering is considered
in terms of a folding of a three-body density functional assuming
separate interactions of electrons with the core and the halo
nucleons. The three-body density functional is obtained from
Faddeev calculations that employ neutron-neutron and neutron-core
forces able to describe the results from collisions with heavy
ions. In \cite{Ant04} various existing theoretical predictions for
the charge distributions in light exotic nuclei $^{6,8}$He,
$^{11}$Li, $^{14}$Be, $^{17,19}$B have been used for calculations
of charge form factors. These were those of Tanihata \textit{et
al.} (\cite{Tan92} for He isotopes), the results of the
cluster-orbital shell-model approximation (COSMA) (\cite{Zhu94}
for He isotopes and \cite{Kor97a} for Li isotopes), the
large-scale shell-model (LSSM) method (\cite{Kar00} for He
isotopes and \cite{Kar97} for Li isotopes) and that of Suzuki
\textit{et al.} \cite{Suz99} for $^{14}$Be and $^{17,19}$B nuclei.
The charge form factors have been calculated within the plane wave
Born approximation. Calculations of form factors of heavier exotic
nuclei within the PWBA are also presented in
Refs.~\cite{Wang04,Richter03}.

The aim of this work is as follows. Firstly, to extend in
comparison with \cite{Ant04} the range of exotic nuclei for which
charge form factors are calculated. Along with the new
calculations for He and Li isotopes, we present results on charge
form factors of several unstable isotopes of medium (Ni) and heavy
(Kr and Sn) nuclei and compare them to those of stable isotopes in
the same isotopic chain. The isotopes of Ni and Sn are chosen
because they have been indicated in Refs.~\cite{Sud01,Kat03} as
first candidates accessible for the charge densities and rms radii
determination and as key isotopes for structure studies of
unstable nuclei at the electron-radioactive-ion collider in RIKEN.
We also give the charge densities and compare them to matter
density distributions. The calculated proton, neutron, charge and
matter rms radii are also presented and the latter are compared
with those for $^{4,6,8}$He and $^{6,11}$Li deduced from the
proton scattering experiments at GSI \cite{Ege02} and from the
total interaction cross sections $\sigma_{I}$
\cite{Tan85a,Tan85b,Tan88a} obtained from the measurements of
Tanihata {\it et al.} \cite{Tan88b,Tan92} and from the re-analysis
\cite{AlKh96,Tost97} of the same data. In our calculations for the
He and Li isotopes we do not use (in contrast to the work of
Ref.~\cite{Ant04}) the semi-phenomenological densities of Tanihata
and COSMA mentioned above, where the parameter values of the
densities were established by a comparison with the total
interaction cross sections. Both densities have unrealistic
Gaussian tails at large $r$. Instead, we use for these nuclei the
LSSM proton and neutron densities obtained in calculations based
on the set of wave functions with exponential asymptotic behaviour
(\cite{Kar00} for He and \cite{Kar97} for Li isotopes). For the
isotopes of heavier nuclei Ni, Kr and Sn we use proton and neutron
densities which are obtained from self-consistent mean-field
(HF+BCS, in short HFB) calculations with density-dependent Skyrme
effective interactions in a large harmonic-oscillator (HO) basis
\cite{SarXX,Sar99}. Secondly, in contrast to the work of
Ref.~\cite{Ant04}, we calculate the charge form factors not only
within the PWBA but also in DWBA by the numerical solution of the
Dirac equation \cite{Yen54,Luk02,Nis85} for electron scattering in
the Coulomb potential of the charge distribution of a given
nucleus. Also, now we do not neglect neutrons, as was done in
Ref.~\cite{Ant04}.

A brief representation of the theoretical scheme is given in
Section~\ref{s:theo}. The results and discussion are given in
Section~\ref{s:res}. The conclusions are summarized in
Section~\ref{s:con}.

\section{The Theoretical Scheme \label{s:theo}}

\subsection{The Form Factors}

In this Section we review briefly the basic formulae used to
calculate the form factors, as well as the proton and neutron
densities.

The nuclear charge form factor $F_{ch}(q)$ has been calculated as
follows
\begin{equation}
F_{ch}(q)= \left [
F_{point,p}(q)G_{Ep}(q)+\frac{N}{Z}F_{point,n}(q)G_{En}(q)\right
]F_{c.m.}(q), \label{eq:1}
\end{equation}
where $F_{point,p}(q)$ and $F_{point,n}(q)$ are the form factors
which are related to the point-like proton and neutron densities
$\rho_{point,p}(\mathbf{r})$ and $\rho_{point,n}(\mathbf{r})$,
respectively. These densities correspond to wave functions in
which the positions ${\bf r}$ of the nucleons are defined with
respect to the centre of the potential related to the laboratory
system. In PWBA these form factors have the form
\begin{equation}
F_{point,p}({\bf q})=\frac{1}{Z}\int \rho_{point,p}({\bf
r})e^{i{\bf q}{\bf r}}d{\bf r}
\label{eq:2}
\end{equation}
and
\begin{equation}
F_{point,n}({\bf q})=\frac{1}{N}\int \rho_{point,n}({\bf
r})e^{i{\bf q}{\bf r}}d{\bf r},
\label{eq:3}
\end{equation}
where
\begin{equation}
\int \rho_{point,p}({\bf r})d{\bf r}=Z; \;\;\;\;  \int
\rho_{point,n}({\bf r})d{\bf r}=N. \label{eq:4}
\end{equation}
In order that $F_{ch}(q)$ corresponds to density distributions in
the centre-of-mass coordinate system, a factor $F_{c.m.}(q)$ is
introduced (e.g. \cite{For66,Alk91,Bur77}) in the standard way
[$F_{c.m.}(q)=exp({q^2/4A^{2/3}})$]. In Eq.~(\ref{eq:1})
$G_{Ep}(q)$ and $G_{En}(q)$ are the Sachs proton and neutron
electric form factors, correspondingly, and they are taken from
one of the most recent phenomenological parametrizations
\cite{Friedrich03}. Actually, there is no significant difference
between this recent parametrization and the most traditional one
of Refs.~\cite{Sim80,Galster71,Chan76} in the range of momentum
transfer considered in this work ($q<4$ fm$^{-1}$).

In the present work, in addition to PWBA, we also perform DWBA
calculations solving the Dirac equation which contains the central
potential arising from the proton ground-state distribution. We
use two codes for the numerical calculations of the form factors:
i) that of \cite{Luk02} which follows Ref.~\cite{Yen54} and ii)
the code from \cite{Nis85}. The results of both calculations were
found in good agreement.

\subsection{The Density Distributions}

The theoretical predictions for the point-like proton and neutron
nuclear densities of the light exotic nuclei $^{6,8}$He and
$^{11}$Li, as well as of the corresponding stable isotopes $^4$He
and $^6$Li are taken from the LSSM calculations. For $^{4,6,8}$He
nuclei they are obtained in a complete $4\hbar\omega$ shell-model
space \cite{Kar00}. The LSSM calculations use a Woods-Saxon
single-particle wave function basis for $^6$He and $^8$He and HO
one for $^4$He. For comparison we use also the ``experimental''
charge density for $^4$He \cite{Bur77} and \cite{Vri87,Sick82},
i.e. the so-called "model-independent" shape of the density. The
proton and neutron densities of $^6$Li are obtained within the
LSSM in a complete $4\hbar\omega$ shell-model space and of
$^{11}$Li in complete $2\hbar\omega$ shell-model calculations
\cite{Kar97}. For $^6$Li the single-particle HO wave functions
have been used in the LSSM calculations and Woods-Saxon ones for
$^{11}$Li. For $^6$Li we also use the point-proton nuclear density
distribution taken from \cite{Pat03,Fri94} which leads to the
``experimental'' charge distribution with rms radius equal to 2.57
fm \cite{Pat03}.

The point proton and neutron density distributions of Ni, Kr and
Sn isotopes are taken from deformed self-consistent HFB
calculations with density-dependent SG2 effective interactions
using a large HO basis with 11 major shells
\cite{Sar99,Vautherin}.

\section{Results and Discussion \label{s:res}}

We calculate charge form factors for a variety of exotic nuclei
with both PWBA and DWBA. As mentioned above, the proton and
neutron densities used for He and Li isotopes are obtained from
realistic microscopic calculations with the LSSM method
\cite{Kar00,Kar97}, while the densities used for Ni, Kr and Sn
isotopes are calculated in the deformed self-consistent HF+BCS
method.

\begin{figure}[htb]
\includegraphics[width=15cm]{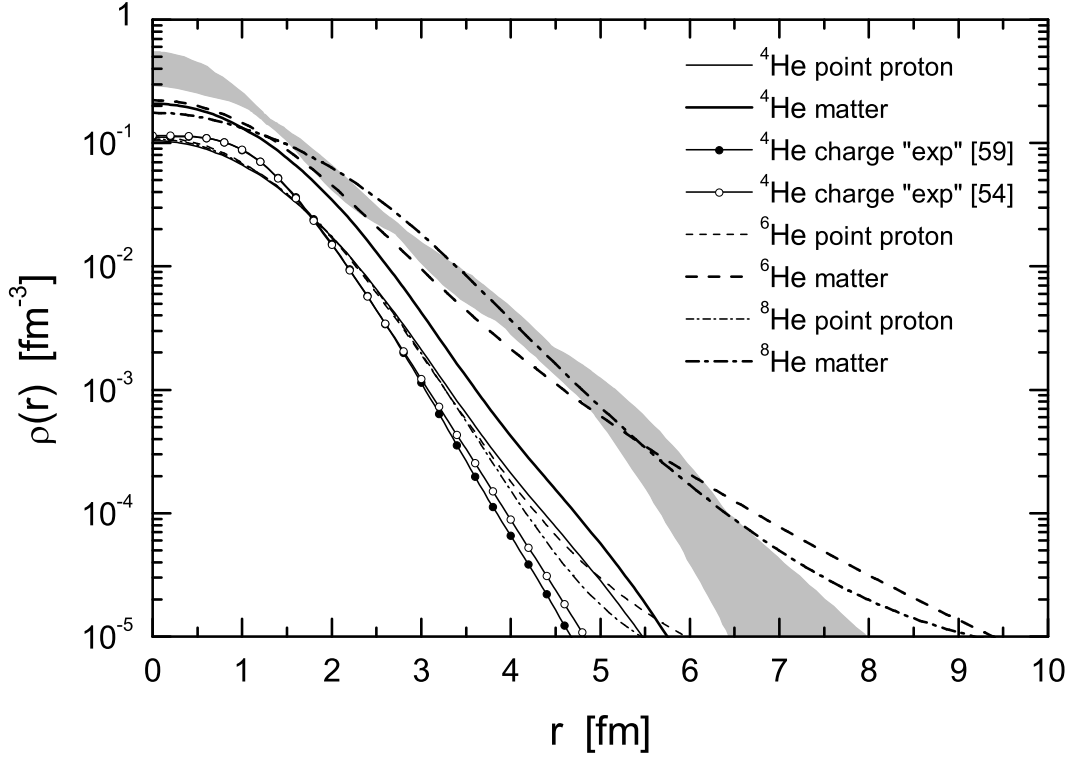}
\caption{Thin lines are LSSM point proton densities of
$^{4,6,8}$He compared to the ``experimental'' charge density for
$^4$He from "model-independent" analyses
\protect\cite{Vri87,Bur77}. Thick lines are LSSM matter densities
of $^{4,6,8}$He compared to matter density of $^{8}$He deduced
from the experimental proton scattering cross section data in
\protect\cite{Ege03} (grey area). \label{fig1}}
\end{figure}

\begin{figure}[htb]
\includegraphics[width=15cm]{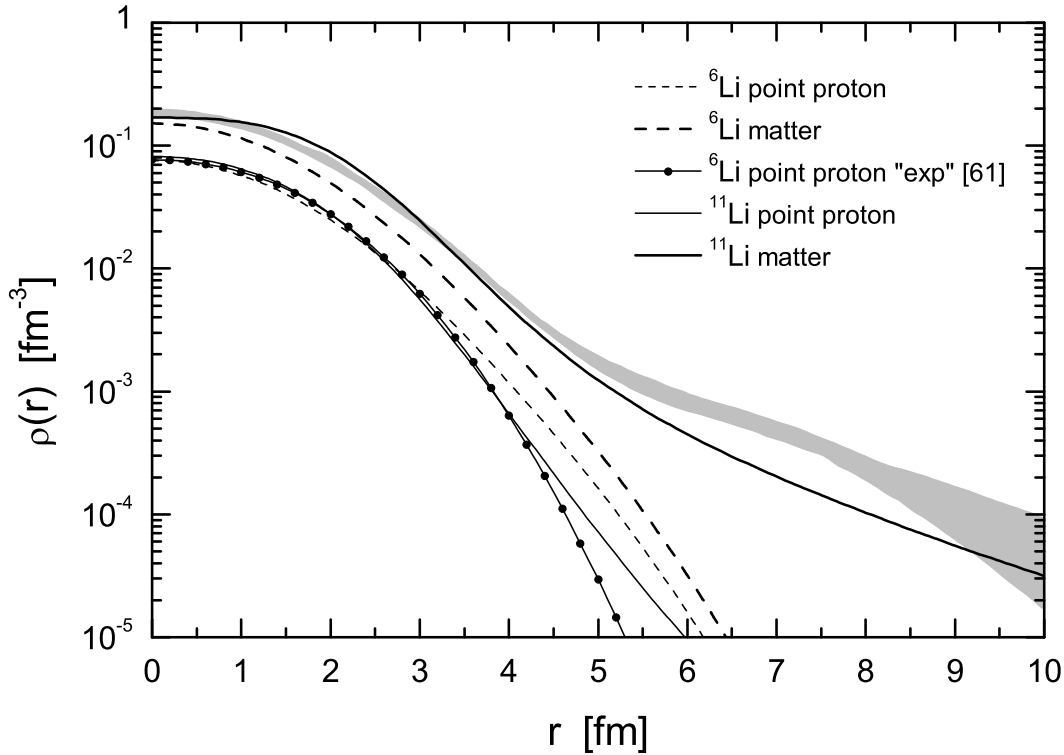}
\caption{Thin lines are LSSM point proton densities of $^{6,11}$Li
compared to the point-proton density of $^6$Li extracted from the
``experimental'' charge density in a "model-independent" analysis
\protect\cite{Pat03}. Thick lines are LSSM matter densities of
$^{6,11}$Li compared to matter density of $^{11}$Li deduced from
the experimental proton scattering cross section data in
\protect\cite{Ege03} (grey area). \label{fig2}}
\end{figure}

Let us first discuss the light nuclei. We show in Figs.~\ref{fig1}
and \ref{fig2} the point proton and matter density distributions
(normalized correspondingly to $Z$ and $A$) calculated with LSSM
for the He isotopes $^{4,6,8}$He \cite{Kar00} and Li isotopes
$^{6,11}$Li \cite{Kar97}. Matter distribution is taken to be
$\rho_m(r) = \rho_{point,p}(r) + \rho_{point,n}(r)$. In addition,
for the sake of completeness of the comparison we give the
``experimental'' charge density of the stable isotope $^4$He (in
Fig.~\ref{fig1}) \cite{Vri87,Bur77} and the point-proton density
of the $^6$Li nucleus (in Fig.~\ref{fig2}) extracted from the
"experimental" charge density in \cite{Pat03}.

Firstly, one can see from Fig.~\ref{fig1} the considerable
difference between the ``experimental'' charge density of $^4$He
and the point proton densities of $^{4,6,8}$He calculated in LSSM
which is also informative of the role of the charge distribution
of the proton itself. Secondly, the differences between the LSSM
proton density of $^4$He and those of $^6$He and $^8$He are not so
large. The only change occurs in the high-$r$ tail, mainly due to
the different (HO versus Woods-Saxon) basis used in the LSSM
calculations of $^4$He. Much more noticeable, however, is the
difference between the LSSM point proton densities in $^{6}$Li and
$^{11}$Li seen in Fig.~\ref{fig2}. There is also a difference at
large values of {\it r} between the LSSM proton density of
$^{6}$Li and the point-proton density of the same nucleus
extracted from the "experimental" charge density in a
"model-independent" analysis \cite{Pat03}. As expected, the matter
distributions of neutron-rich $^{6,8}$He and $^{11}$Li are quite
different from those of the stable $^4$He and $^6$Li both in the
surface region and in the interior of nuclei.

For comparison we present by grey area also the matter densities
of $^{8}$He (in Fig.~\ref{fig1}) and $^{11}$Li (in
Fig.~\ref{fig2}) deduced from the experimental data for the
differential cross sections of elastic proton scattering at small
momentum transfer which have been measured at GSI at energies
around 700 MeV/nucleon in inverse kinematics for neutron-rich
helium and lithium isotopes \cite{Ege03}. A model-dependent method
to extract the matter distributions was used for these nuclei
exploring various parametrizations for the nucleon density
distributions. The calculated LSSM matter distribution for $^8$He
is in agreement with that extracted from proton scattering data
\cite{Ege03} in the interval $2\leq r \leq 7$ fm. For $^{11}$Li
this is the case in the interval $0\leq r \leq 4$ fm.

\begin{figure}[htb]
\includegraphics[width=17cm]{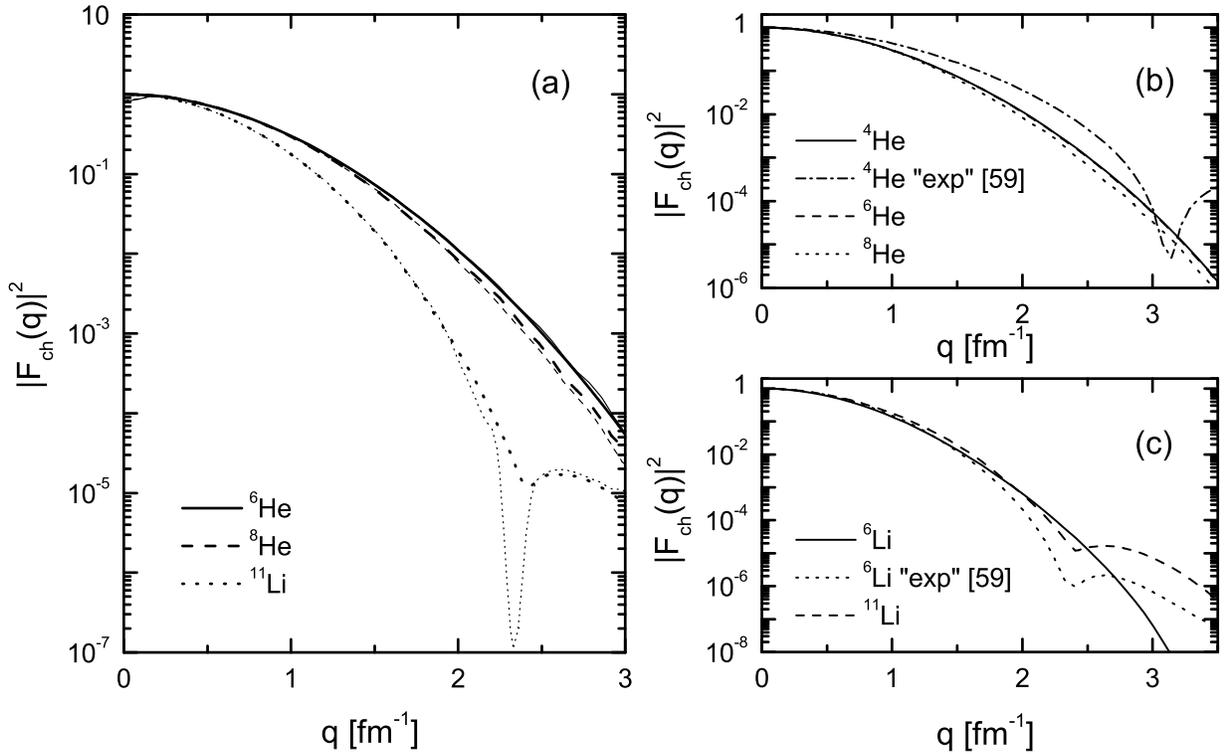}
\caption{(a) Charge form factors of $^6$He, $^8$He and $^{11}$Li
calculated in PWBA (thin lines) and in DWBA (thick lines) using
LSSM densities; (b) charge form factors in DWBA for $^4$He
(calculated by using ``experimental'' charge density
\protect\cite{Vri87} and the LSSM density) and of $^{6,8}$He
(using the LSSM densities); (c) charge form factor in DWBA for
$^6$Li (using the ``experimental'' charge density
\protect\cite{Vri87} and the LSSM densities) and for $^{11}$Li
(using the LSSM densities). \label{fig3}}
\end{figure}

In Fig.~\ref{fig3}(a) the results for the charge form factors [Eq.
(\ref{eq:1})] of $^{6,8}$He and $^{11}$Li obtained in PWBA (thin
lines) and in DWBA (thick lines) using LSSM densities are shown.
In Fig.~\ref{fig3}(b) the charge form factors of $^4$He obtained
in DWBA by means of the ``experimental'' \cite{Vri87} and LSSM
charge density are compared with those of $^{6}$He and $^{8}$He.
The same is shown in Fig.~\ref{fig3}(c) for the $^6$Li nucleus in
comparison with the form factor of $^{11}$Li (using its LSSM
densities). The DWBA calculations are performed at an energy of
540 MeV. One can see from Fig.~\ref{fig3}(a) the small difference
of the charge form factors of $^{6}$He and $^{8}$He at $q\geq $ 1
fm$^{-1}$ and the small deviation of the DWBA from PWBA results in
the whole $q$-range. It is shown in Fig.~\ref{fig3}(b) the similarity
of the LSSM charge form factors of $^{4}$He and $^{6}$He and their
difference from that of $^{8}$He. At the same time there is not a
minimum in this
$q$-range in all three LSSM form factors of $^{4,6,8}$He in
contrast to the case for the ``experimental'' charge form factor
of $^{4}$He.

\begin{figure}[htb]
\includegraphics[width=17cm]{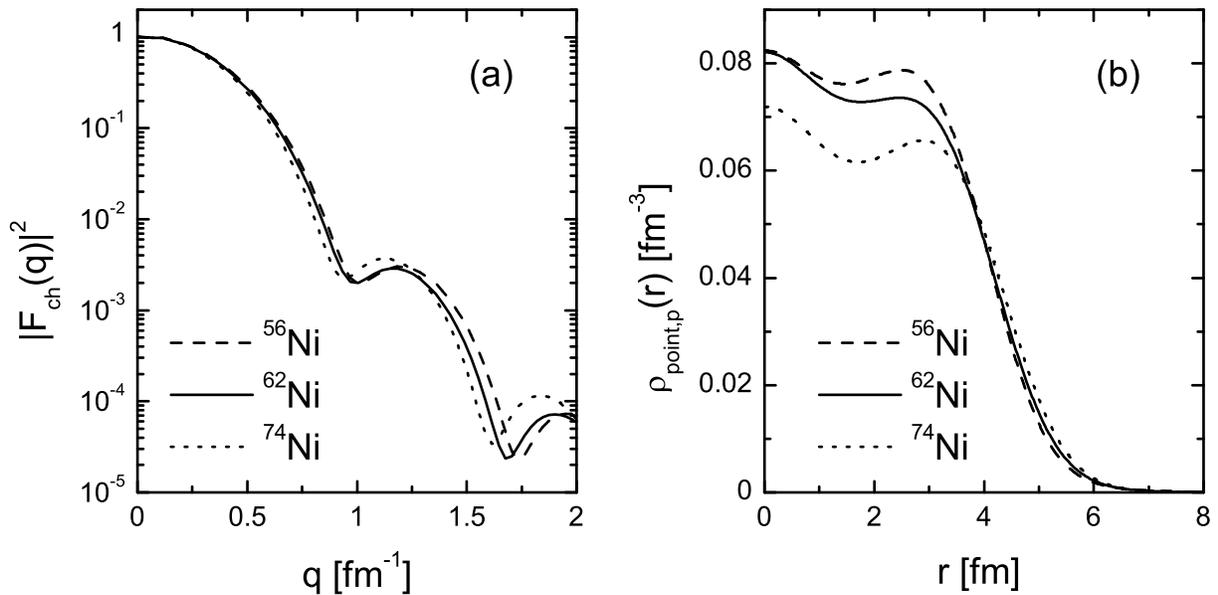}
\caption{(a) Charge form factors for the unstable doubly-magic
$^{56}$Ni, stable $^{62}$Ni and unstable $^{74}$Ni isotopes
calculated by using the HF+BCS densities and the DWBA; (b) HF+BCS
proton densities of $^{56}$Ni, $^{62}$Ni and $^{74}$Ni.
\label{fig4}}
\end{figure}

\begin{figure}[htb]
\includegraphics[width=17cm]{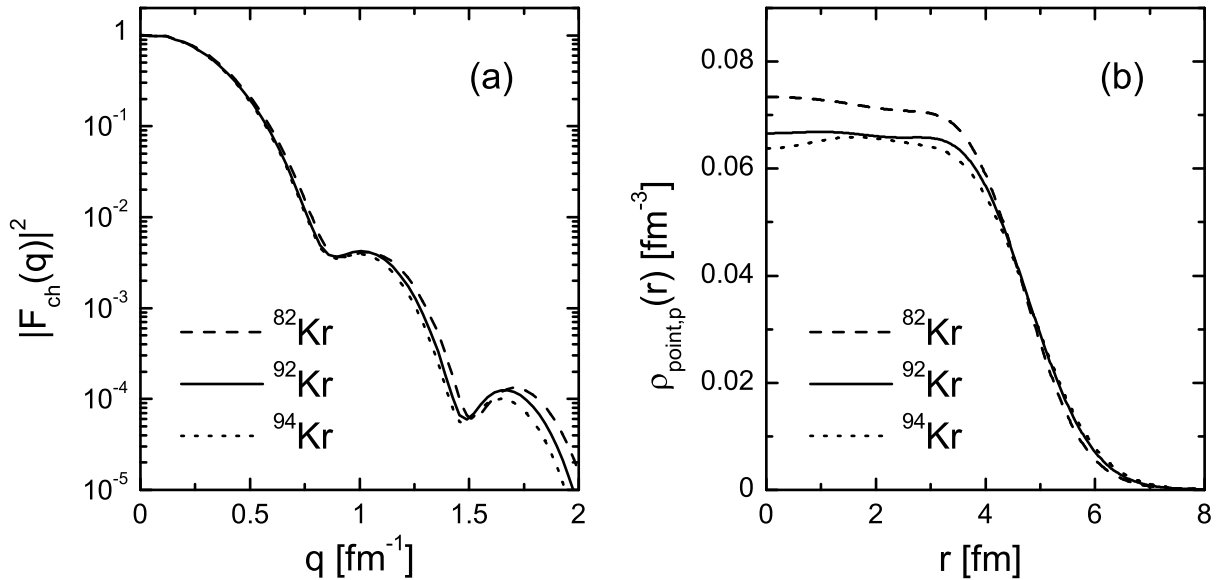}
\caption{(a) Charge form factors for the stable isotope $^{82}$Kr
and for the unstable $^{92}$Kr and $^{94}$Kr isotopes calculated
by using the HF+BCS densities and the DWBA; (b) HF+BCS proton
densities of $^{82}$Kr, $^{92}$Kr and $^{94}$Kr. \label{fig5}}
\end{figure}

\begin{figure}[htb]
\includegraphics[width=17cm]{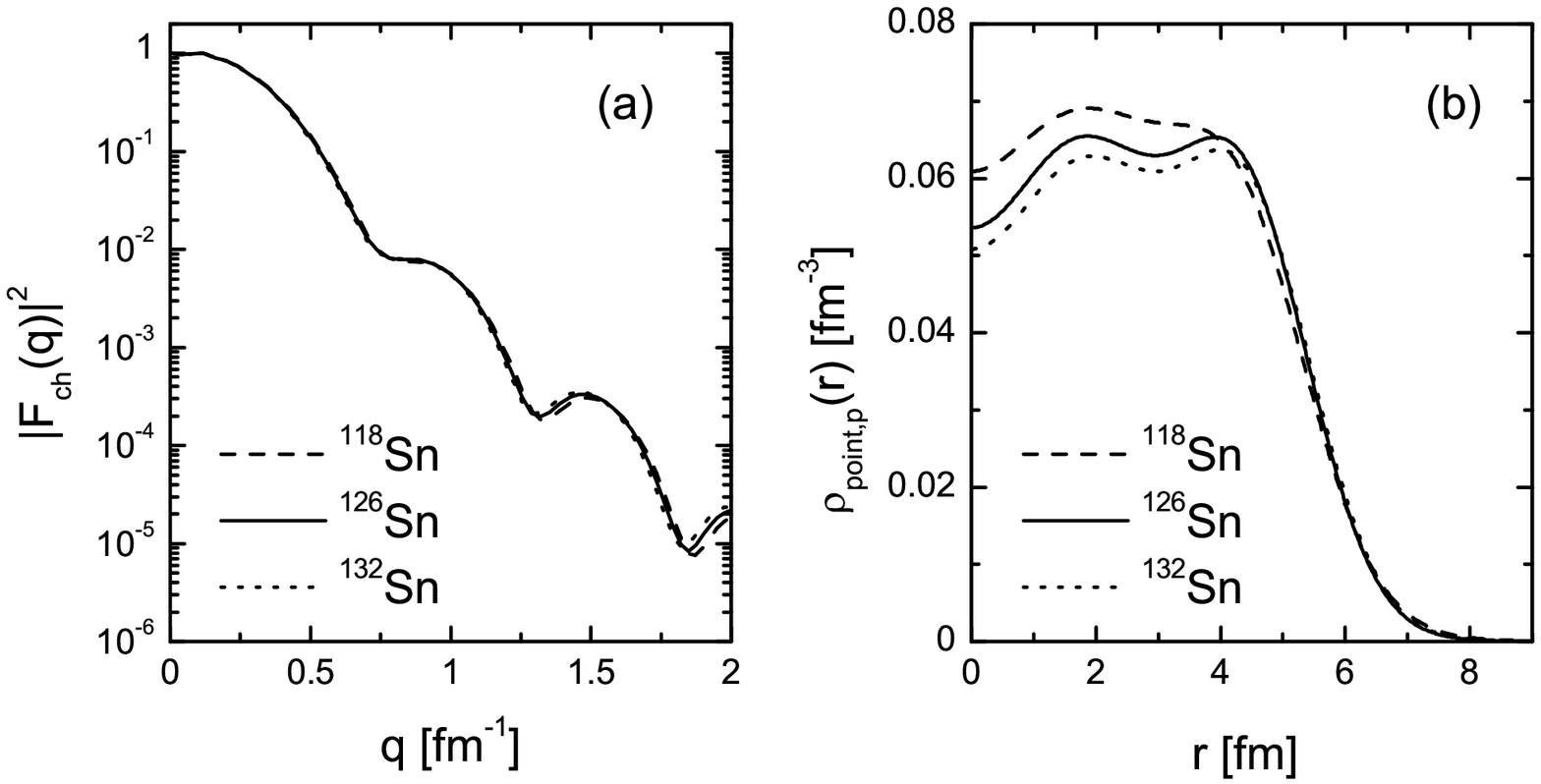}
\caption{(a) Charge form factors for the stable isotope
$^{118}$Sn, unstable $^{126}$Sn and unstable doubly-magic
$^{132}$Sn isotopes calculated by using the HF+BCS densities and
the DWBA; (b) HF+BCS proton densities of $^{118}$Sn, $^{126}$Sn
and $^{132}$Sn. \label{fig6}}
\end{figure}

In Figs.~\ref{fig4}, \ref{fig5} and \ref{fig6} we present the
charge form factors calculated with DWBA at an energy of 250 MeV
as well as the HF+BCS proton densities for $^{56,62,74}$Ni,
$^{82,92,94}$Kr and $^{118,126,132}$Sn, correspondingly. A common
feature of the charge form factors of the Ni, Kr and Sn isotopes
considered, which can be seen in Figs.~\ref{fig4}(a),
\ref{fig5}(a) and \ref{fig6}(a), is the shift of the minima to
smaller values of $q$ when the number of neutrons increases in a
given isotopic chain. This is due mainly to the enhancement of the
proton densities in the peripherical  region and also (to a minor
extent) to the contribution of the charge distribution of the
neutrons themselves. Indeed, one can see from Figs.~\ref{fig4}(b),
\ref{fig5}(b) and \ref{fig6}(b) that the point proton densities in
a given isotopic chain decrease in the central region and increase
in the surface with increasing neutron number.

The isotopic sensitivities of the calculated charge form factors
to the changes of neutron number observed in Figs.~\ref{fig4}(a),
\ref{fig5}(a) and \ref{fig6}(a) and their precise measuring in
future electron-nucleus scattering experiments may lead to
accurate determination of charge distributions for unstable
nuclei. The techniques used to extract charge distributions from
the measured elastic form factors are well established. For
instance, the model-dependent method in which the direct
scattering problem is solved parametrizing the charge distribution
and the respective parameters are fitted to the experimental cross
sections, has been demonstrated in \cite{GSI2005} to show the
sensitivity of the cross section  (and of the charge form factor,
respectively) to variations in radius and diffuseness parameters.
The data were simulated for the $^{132}$Sn$(e,e)$ elastic
scattering at a luminosity of 10$^{28}$ cm$^{-2}$s$^{-1}$. This
model estimation shows that at low-momentum transfers ($q<1.5$
fm$^{-1}$) the charge form factor of $^{132}$Sn can be precisely
measured. However, in the range of moderate- and high-momentum
transfer, where the charge form factor is dominated by the details
of the charge density distribution, the expected error band
becomes appreciable. Hence, covering a wider region of $q$ makes
possible to determine the charge distribution but requires higher
luminosities. There is a qualitative agreement of model-dependent
calculations of charge form factors for Sn nucleus in \cite{Sud01}
with our results shown in Fig.~\ref{fig6}(a). Another way to
extract the charge distributions is to use a model-independent
analysis based upon the expansion of the charge density on a
complete set of orthogonal functions. Such type of analysis allows
one to show whether the isotopic effects on charge densities can
be measured convincingly. Since the charge distribution of
unstable nuclei is the main subject of the coming experiments at
next-generation electron-nucleus  colliders, this problem deserves
further study.

\begin{figure}[htb]
\includegraphics[width=12cm]{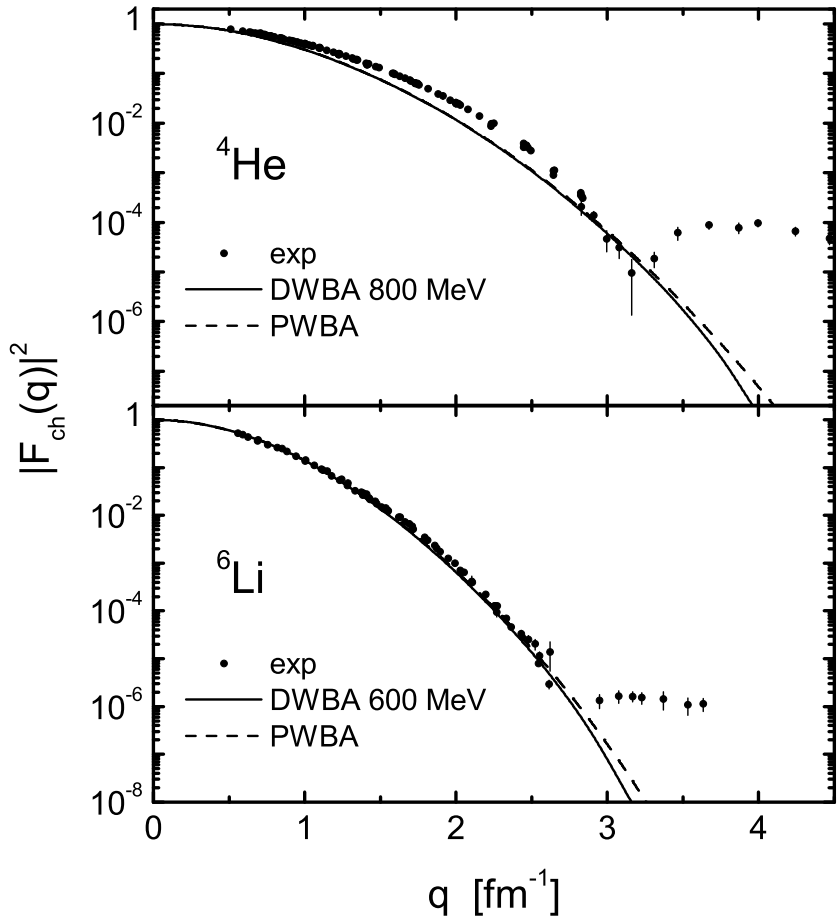}
\caption{Charge form factors for the stable isotopes $^{4}$He and
$^{6}$Li calculated using LSSM densities in PWBA and in DWBA in
comparison with the experimental data.} \label{fig7}
\end{figure}

\begin{figure}[htb]
\includegraphics[width=12cm]{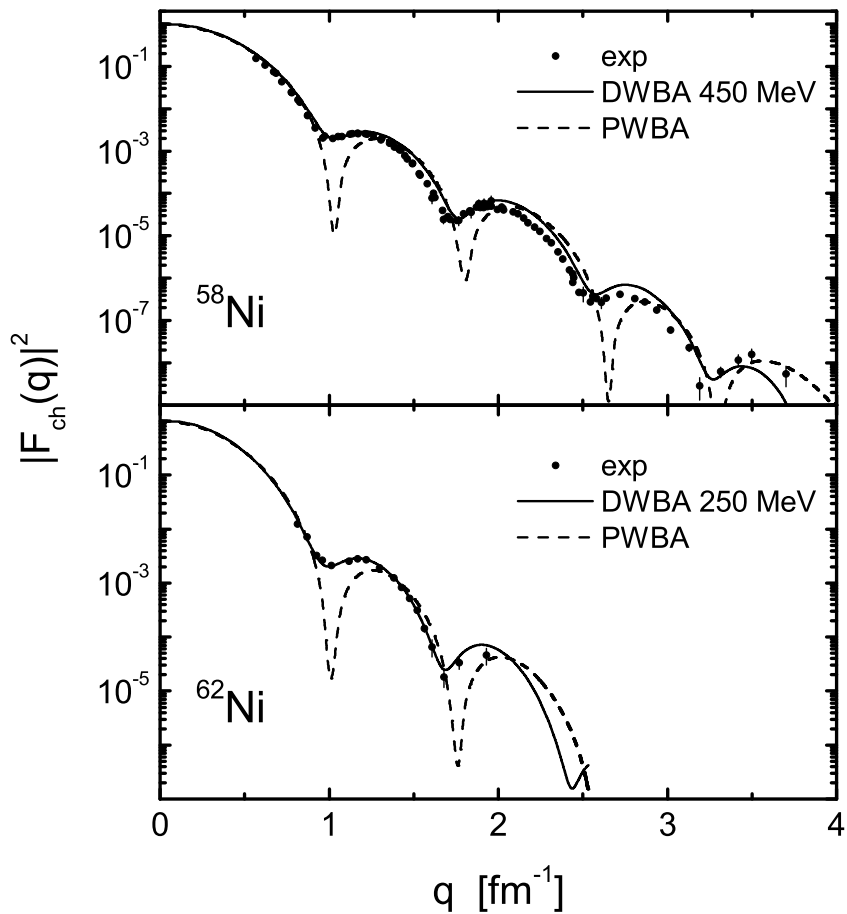}
\caption{Charge form factors for the stable isotopes $^{58}$Ni and
$^{62}$Ni calculated by using the HF+BCS densities and the PWBA
and DWBA in comparison with the experimental data.}
\label{fig8}
\end{figure}

\begin{figure}[htb]
\includegraphics[width=12cm]{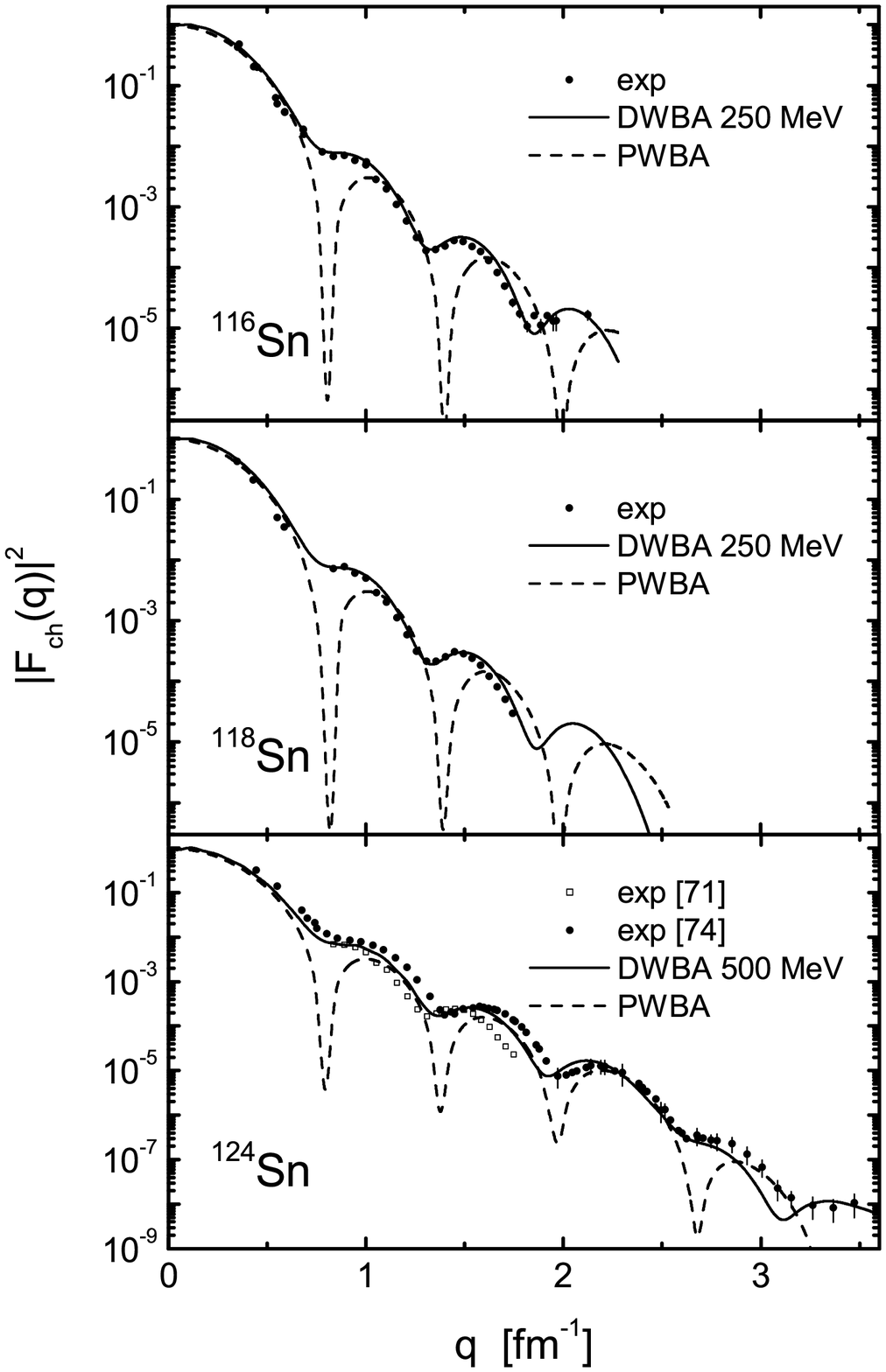}
\caption{Charge form factors for the stable isotopes $^{116}$Sn,
$^{118}$Sn and $^{124}$Sn calculated by using the HF+BCS densities
and the PWBA and DWBA in comparison with the experimental data.}
\label{fig9}
\end{figure}

For the sake of completeness we show the comparison of the DWBA
results with available experimental data for the charge form
factors of the isotopes $^{4}$He \cite{McCarthy77,Otter85} and
$^{6}$Li \cite{Suelzle67,Li71} (Fig.~\ref{fig7}), $^{58}$Ni
\cite{Sick75} and $^{62}$Ni \cite{Lit71} (Fig.~\ref{fig8}),
$^{116}$Sn \cite{Lit72,Curtis69,Light76}, $^{118}$Sn
\cite{Lit72,Curtis69} and $^{124}$Sn \cite{Lit72,Cavedon82}
(Fig.~\ref{fig9}). Our DWBA calculations are performed at the
electron energies used in the experiments. The agreement with the
empirical data for the stable isotopes is supportive of our
results on the exotic nuclei to be used as guidance to future
experiments particularly so on the medium-heavy and heavy ones. A
common feature is the expected filling of the Born zeros when DWBA
is used (instead of PWBA), as well as the shift of the minima to
smaller values of $q$ and the increase of the secondary peaks
which can be seen in Figs.~\ref{fig8} and \ref{fig9}.

\begin{figure}[htb]
\includegraphics[width=17cm]{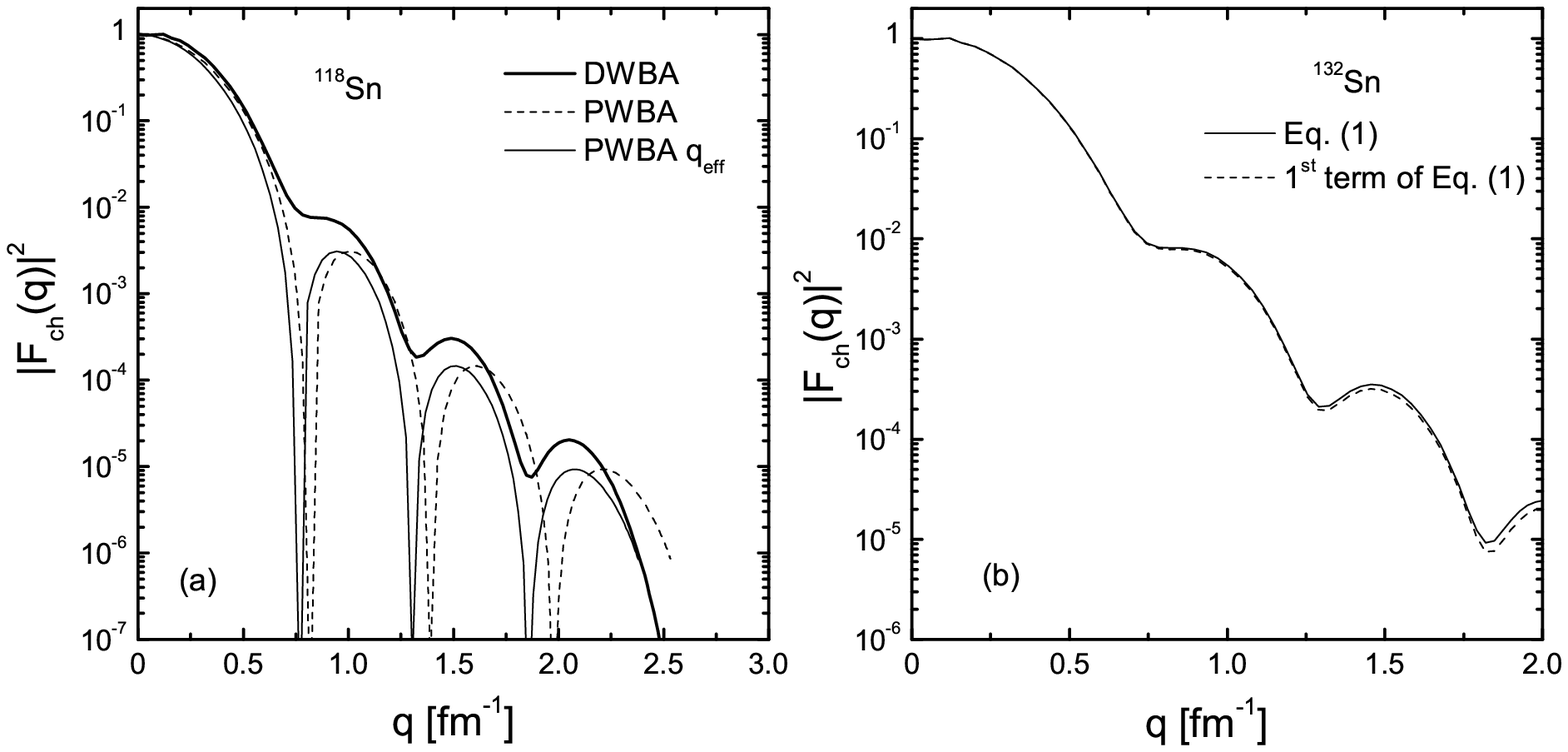}
\caption{(a) Charge form factors for the stable isotope $^{118}$Sn
calculated by using the DWBA (thick solid line), PWBA (dashed
line) and PWIA with $q_{eff}$ given in the text (thin solid line);
(b) Charge form factor for the unstable doubly-magic $^{132}$Sn
isotope calculated by using the DWBA and corresponding to
Eq.~(\ref{eq:1}) (solid line) and to the proton contribution only
[i.e. to the first term of Eq.~(\ref{eq:1})] (dashed line).}
\label{fig10}
\end{figure}

In this spirit we would also like to note that the displacement to
the left of the DWBA calculations versus PWBA can be accounted for
by replacement of the momentum transfer $q$ with the effective
momentum transfer $q_{eff}$ (see, e.g., \cite{Ube71}). We take
into account this correction (which is due to the Coulomb
attraction felt by the electrons) by using $q_{eff} = q [
1+(cZ\alpha / R_{ch}E_i)]$, where the constant $c$ (in our work
$c=1$) is related to the charge rms radii $R_{ch}$ obtained in the
present calculations. The effect of using $q_{eff}$ is clearly
seen in Fig.~\ref{fig10}(a) on the example of $^{118}$Sn isotope.
It describes the shift of the minima produced by the Coulomb
distortion of the electron waves. To illustrate the effect of the
neutron form factor on the nuclear charge form factor, we show in
Fig.~\ref{fig10} (b) for the case of $^{132}$Sn the results
corresponding to the total charge form factor $F_{ch}(q)$ as
defined in Eq.~(\ref{eq:1}) and to its proton contribution
$F_{point,p}(q)G_{Ep}(q)$. As can be seen from the Figure,
although the contribution from the neutrons is rather small
(around 10-20\% in the $q$-range $1.5\div 2$ fm$^{-1}$), it is
comparable in size to the isotopic effect and, therefore, should
not be neglected.

We would like to note the reasonable agreement of the results of
the DWBA calculations with the experimental charge form factors of
the isotopes of Ni and Sn considered. The lack of theoretical
minima for $^{4}$He and $^{6}$Li, however, leads us to the
conclusion that the LSSM densities of these light stable isotopes
do not seem reliable. The latter might be due to the use of
harmonic-oscillator wave functions in the LSSM calculations for these
nuclei.

In Tables \ref{table1} and \ref{table2} we give the rms radii
($R_{p}$, $R_{n}$, $R_{ch}$, $R_{m}$) corresponding to nuclear
proton, neutron, charge and matter distributions, as well as the
difference $\Delta R=R_{m}-R_{p}$ for the He and Li isotopes
(Table \ref{table1}) and for the Ni, Kr and Sn isotopes (Table
\ref{table2}) which are considered in our work. The values of the
diffuseness parameter of the various densities are presented in
Table \ref{table3}. The diffuseness parameter is defined as the
distance over which the value of the density decreases in the
surface region from 90 \% to 10 \% of its value in the centre of
the nucleus divided by 4.4. For comparison we give additionally in
Table \ref{table1} the nuclear matter radii of $^4$He, $^6$He,
$^8$He and $^6$Li, $^{11}$Li deduced from the proton scattering
experiments at GSI \cite{Ege02}, from the data on total
interaction cross sections $\sigma_{I}$
\cite{Tan85a,Tan85b,Tan88a} obtained from an analysis of Tanihata
{\it et al.} \cite{Tan92,Tan88b} and from a more recent
re-analysis \cite{AlKh96,Tost97} of the same data. We present in
Table \ref{table1} for a comparison also the experimental charge
rms radii for $^4$He and $^6$Li from \cite{Bur77,Vri87,Pat03} and
in Table \ref{table2} those for $^{58,62}$Ni and $^{116,118}$Sn
from \cite{Pat03} and for $^{82,92,94}$Kr from \cite{Keim95}.

\begin{table}[h]
\caption{Proton ($R_{p}$), neutron ($R_{n}$), charge ($R_{ch}$),
matter ($R_{m}$) rms radii (in fm) and difference $\Delta
R=R_{m}-R_{p}$ of He and Li isotopes calculated using LSSM
densities. Available data on $R_{m}$ and $R_{ch}$ are also
presented.}
{\begin{tabular}{ccccccccccccc} \hline \hline Nuclei &
& $R_{p}$ & $R_{n}$ & $R_{ch}$ & $R_{m}$ & $\Delta R$ & $R_{m}$
\protect\cite{Ege02} & $R_{m}$ \protect\cite{Tan92,Tan88b} &
$R_{m}$ \protect\cite{AlKh96,Tost97} &
$R_{ch}$ \protect\cite{Vri87,Pat03}& $R_{ch}$ \protect\cite{Bur77} \\
\hline
$^{4}$He    & & 1.927 & 1.927 & 2.153 & 1.927 & 0.000 & 1.49(3) &          &          & 1.696(14) & 1.695 \\
$^{6}$He    & & 1.945 & 2.900 & 2.147 & 2.621 & 0.676 & 2.30(7) & 2.33(4)  & 2.54(4)  &           &       \\
$^{8}$He    & & 1.924 & 2.876 & 2.140 & 2.670 & 0.746 & 2.45(7) & 2.49(4)  &          &           &       \\
$^{6}$Li    & & 2.431 & 2.431 & 2.647 & 2.431 & 0.000 & 2.45(7) & 2.32(3)  &          & 2.57(10)  & 2.539 \\
$^{11}$Li   & & 2.238 & 3.169 & 2.477 & 2.945 & 0.707 & 3.62(19)& 3.12(16) & 3.53(10) &           &       \\
\hline \hline
\end{tabular}}
\label{table1}
\end{table}

\begin{table}[h]
\caption{Proton ($R_{p}$), neutron ($R_{n}$), charge ($R_{ch}$),
matter ($R_{m}$) rms radii (in fm) and difference $\Delta
R=R_{m}-R_{p}$ of Ni, Kr and Sn isotopes calculated using HF+BCS
densities. The last two columns present experimental data on
$R_{ch}$.}
{\begin{tabular}{cccccccccc} \hline \hline Nuclei & &
$R_{p}$ & $R_{n}$ & $R_{ch}$ & $R_{m}$ & $\Delta R$ & $R_{ch}$
\protect\cite{Pat03}
& $R_{ch}$ \protect\cite{Keim95} \\
\hline
$^{56}$Ni    & & 3.725 & 3.666 & 3.795 & 3.696 & -0.029 &           &           \\
$^{58}$Ni    & & 3.719 & 3.697 & 3.794 & 3.707 & -0.012 & 3.764(10) &           \\
$^{62}$Ni    & & 3.798 & 3.855 & 3.866 & 3.829 &  0.031 & 3.830(13) &           \\
$^{74}$Ni    & & 3.911 & 4.130 & 3.977 & 4.049 &  0.138 &           &           \\
$^{82}$Kr    & & 4.126 & 4.190 & 4.189 & 4.162 &  0.036 &           & 4.192(4)  \\
$^{92}$Kr    & & 4.224 & 4.412 & 4.285 & 4.340 &  0.116 &           & 4.273(16) \\
$^{94}$Kr    & & 4.277 & 4.496 & 4.338 & 4.413 &  0.136 &           & 4.300(20) \\
$^{116}$Sn   & & 4.583 & 4.650 & 4.646 & 4.621 &  0.038 & 4.626(15) &           \\
$^{118}$Sn   & & 4.649 & 4.739 & 4.705 & 4.701 &  0.052 & 4.679(16) &           \\
$^{126}$Sn   & & 4.642 & 4.798 & 4.698 & 4.737 &  0.095 &           &           \\
$^{132}$Sn   & & 4.685 & 4.879 & 4.740 & 4.807 &  0.122 &           &           \\
\hline \hline
\end{tabular}}
\label{table2}
\end{table}

\begin{table}[h]
\caption{Diffuseness parameter values (in fm) of the LSSM
densities of He and Li isotopes and HF+BCS densities of Ni, Kr and
Sn isotopes considered in this work.} {\begin{tabular}{ccccccc}
\hline \hline
Nuclei & & $a_{p}$ & $a_{n}$ & $a_{m}$ & $a_{ch}$ \\
\hline
$^{4}$He     & & 0.407 & 0.407 & 0.407 & 0.392 \\
$^{6}$He     & & 0.397 & 0.498 & 0.448 & 0.381 \\
$^{8}$He     & & 0.403 & 0.513 & 0.549 & 0.387 \\
$^{6}$Li     & & 0.521 & 0.521 & 0.521 & 0.509 \\
$^{11}$Li    & & 0.482 & 0.444 & 0.493 & 0.478 \\
$^{56}$Ni    & & 0.484 & 0.505 & 0.493 & 0.527 \\
$^{62}$Ni    & & 0.920 & 0.557 & 0.572 & 0.616 \\
$^{74}$Ni    & & 0.538 & 0.445 & 0.475 & 0.552 \\
$^{82}$Kr    & & 0.509 & 0.459 & 0.477 & 0.570 \\
$^{92}$Kr    & & 0.505 & 0.541 & 0.527 & 0.564 \\
$^{94}$Kr    & & 0.516 & 0.761 & 0.639 & 0.582 \\
$^{118}$Sn   & & 0.468 & 0.555 & 0.509 & 0.534 \\
$^{126}$Sn   & & 0.382 & 0.707 & 0.482 & 0.445 \\
$^{132}$Sn   & & 0.377 & 0.698 & 0.473 & 0.434 \\
\hline \hline
\end{tabular}}
\label{table3}
\end{table}

It is seen from Table \ref{table1} that the calculated rms radii
of He and Li isotopes follow the behaviour of the density
distributions shown in Figs.~\ref{fig1} and \ref{fig2}. One can
also see that the calculated charge rms radii of $^{4}$He and
$^{6}$Li are larger than the experimental ones as could have been
foreseen from Figs.~\ref{fig3}(a) and \ref{fig3}(c). The matter
density of $^{11}$Li exhibits the most extended halo component
(see Fig.~\ref{fig2}) among all helium and lithium isotopes being
investigated, which is reflected in large neutron radius
$R_{n}=3.169$ fm and, correspondingly, in large matter radius
$R_{m}=2.945$ fm. The nuclei $^{6}$He and $^{8}$He have less
extended nuclear matter distributions than $^{11}$Li (see
Fig.~\ref{fig1}) and thus smaller matter radii, $R_{m}=2.621$ fm
and $R_{m}=2.670$ fm compared to 2.945 fm in $^{11}$Li. Our
theoretically calculated $R_{m}$ for $^{6}$He is in closer
agreement with the value from the re-analysis of the data deduced
from total interaction cross sections \cite{Tan88b} performed by
Al-Khalili {\it et al.} \cite{AlKh96}. As for $^{6}$Li, the result
from the present calculation exceeds the value $R_{m}=2.32 (3)$ fm
from Tanihata {\it et al.} \cite{Tan88b}, but almost coincides
with the value $R_{m}=2.45 (7)$ fm deduced from the recent proton
scattering experiments at GSI \cite{Ege02}.

The common tendency of all predicted rms radii for medium (Ni) and
heavy (Kr and Sn) nuclei presented in Table \ref{table2} is the
small increase of their values with the increase of the number of
neutrons in a given isotopic chain except that $R_{ch}$ of
$^{126}$Sn is practically the same as $R_{ch}$ of $^{118}$Sn. Our
theoretical results on $R_{ch}$ in Table~\ref{table2} are in good
agreement with the available experimental values \cite{Pat03,
Keim95}. A more detailed study of the rms radii of these nuclei is
required when future experiments will be performed. In particular,
the charge rms radius can be determined using the
model-independent relation of the form factor in the lower
$q$-region (e.g. \cite{Abbott2000}),
\begin{equation}
R^2_{ch}= -6 \left[ \frac{d F_{ch}(q^2)}{d (q^2)} \right]_{q^2=0}.
\label{eq:5}
\end{equation}

In our opinion, the calculated difference $\Delta R=R_{m}-R_{p}$
whose values are listed also in Tables \ref{table1} and
\ref{table2} is of particular importance and together with the
neutron thickness $R_{n}-R_{p}$ presented in
Refs.~\cite{Ozawa2001,Swia05} can serve as a measure of the halo
or neutron skin structure of neutron-rich exotic nuclei.

\begin{figure}[htb]
\includegraphics[width=17cm]{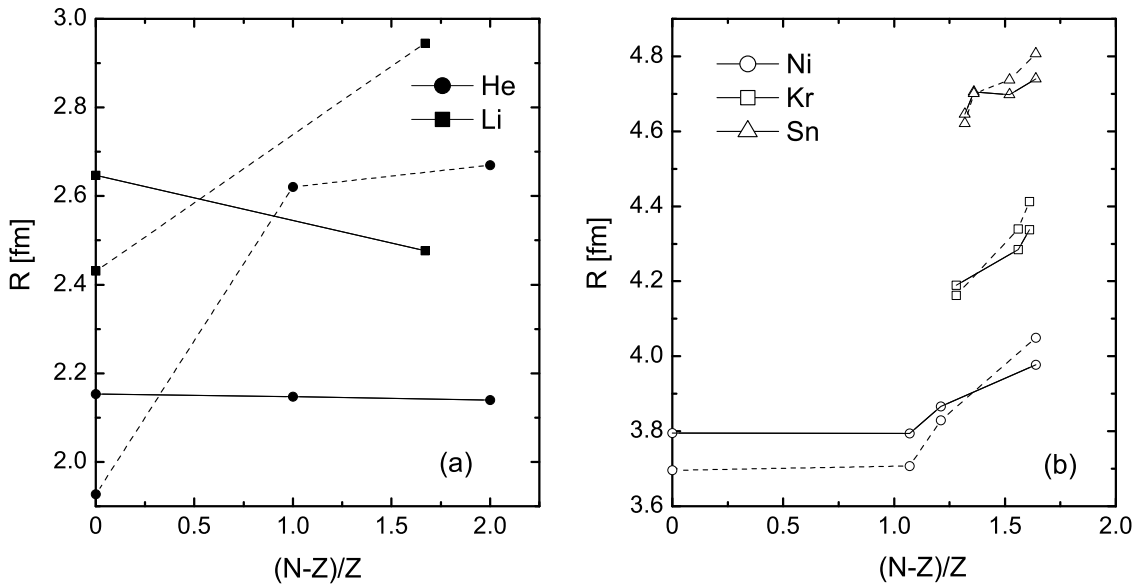}
\caption{Charge ($R_{ch}$) (solid eye-guide lines) and matter
($R_{m}$) (dashed lines) rms radii calculated in this work as a
function of the relative neutron excess $(N-Z)/Z$ of He and Li
isotopes (full symbols) (a) and Ni, Kr and Sn isotopes (open
symbols) (b).}
\label{fig11}
\end{figure}

In addition, we show in Fig.~\ref{fig11} the variation of the
charge and matter rms radii with the relative neutron excess for
all isotopic chains considered. The use of LSSM charge densities
for He and Li isotopes \cite{Kar00,Kar97} leads to a small
decrease of the charge rms radius $R_{ch}$ from $^{4}$He to
$^{6}$He and $^{8}$He and to a larger decrease of $R_{ch}$ from
$^{6}$Li to $^{11}$Li. On the contrary, the behavior of the charge
radii for heavier Ni, Kr and Sn isotopes shows a smooth increase
of $R_{ch}$ with increase of the neutron number, while the nuclear
matter radii for these isotopes increase faster. In order to test
the theoretical predictions for the charge and matter radii, it is
desirable to measure both matter and charge distributions for the
same nuclei. The difference in size of these distributions will be
of high interest and importance for the theoretical understanding
of the exotic nuclei structure.

\section{conclusions \label{s:con}}

The results of the present work can be summarized as follows:

In this work we extended the studies of the previous one
\cite{Ant04} of the proton, neutron, charge and matter densities
and related charge form factors from the light neutron-rich exotic
nuclei $^{6,8}$He, $^{11}$Li to examples of unstable medium (Ni)
and heavy (Kr and Sn) isotopes in comparison with those of stable
isotopes in the same isotopic chain. For He and Li isotopes we use
the proton and neutron densities obtained from realistic
microscopic calculations within the large-scale shell-model method
\cite{Kar00,Kar97}. The densities of Ni, Kr and Sn isotopes are
calculated in HF+BCS method with a density-dependent effective
interaction using a large harmonic-oscillator basis
\cite{SarXX,Sar99}.

We also compare proton and matter density distributions for He and
Li isotopes. The calculated matter distributions for the halo
nuclei are much more extended than the proton ones. We compare
proton density distributions for the isotopes of He, Li, Ni, Kr
and Sn and establish the differences of the proton densities in a
given isotopic chain due to the presence of the neutron excess.
There is a decrease of the proton density in the nuclear interior
and an increase of its tail at large $r$ with increasing neutron
number.

A comparison of the proton, neutron, charge and matter rms radii
as well as the corresponding diffuseness is performed for all
isotopic chains considered. We point out that the general trend of
the difference $\Delta R$ between the matter and proton rms radii
is to increase with the number of neutrons but for the heavy
isotopes this increase is moderate compared to that of the light
ones.

The calculated matter densities for $^8$He and $^{11}$Li are in
fair agreement with the experimental data obtained in proton
scattering on these isotopes in GSI \cite{Ege02}. We compare the
matter rms radii with those from \cite{Ege02} as well as with
those from total interaction cross section data
\cite{Tan85a,Tan85b,Tan88a,Tan92} and their re-analysis
\cite{AlKh96,Tost97}.

We calculate the charge form factors of He, Li, Ni, Kr and Sn
isotopes by means of the densities mentioned above. The charge
form factors are calculated not only in the PWBA as in our
previous work \cite{Ant04} but also in the DWBA, solving the Dirac
equation for electron scattering in the Coulomb potential of the
charge distribution in a given nucleus. By accounting for the
Coulomb distortion of the electron waves the Born zeroes are
filled and the form factors are shifted to smaller values of $q$
which is clearly seen in the cases of the Ni, Kr and Sn isotopes
where $Z$ is large enough. We find that this shift is best
parametrized by $q_{eff} = q [ 1+(Z\alpha / R_{ch}E_i)]$, where
$R_{ch}$ are the charge rms radii as given in the Tables. In
addition we also take into account the charge distribution in the
neutron itself. We find that the contributions from the neutrons
to the charge form factors are less than 20 \% up to $q\sim $ 2
fm$^{-1}$.

The differences between the charge form factors in a given
isotopic chain are shown. The common feature of the charge form
factors is the shift of the form factor curves and their minima to
smaller values of $q$ with the increase of the neutron number in a
given isotopic chain.This is due to the corresponding enhancement
of the proton tails in the peripherical region of the nuclei.

The performed theoretical analyses of the densities and charge
form factors can be a step in the studies of the influence of the
increasing neutron number on the proton and charge distributions
in a given isotopic chain. This is important for understanding the
neutron-proton interaction in the nuclear medium. We  emphasize also
the questions of interest, namely, the necessary both kinematical
regions of the proposed experiments and precision to measure small
shifts in the form factors.

The theoretical predictions for the charge form factors of exotic
nuclei are a challenge for their measurements in the future
experiments in GSI and RIKEN and thus, for obtaining detailed
information on the charge distributions of such nuclei. The
comparison of the calculated charge form factors with the future
data will be a test of the corresponding theoretical models used
for studies of the exotic nuclei structure.

\acknowledgments The authors are grateful to Professor H. Rebel
for the discussion and to Dr. S. Karataglidis for providing us
with the results on the LSSM densities for helium and lithium
isotopes. Three of the authors (A.N.A., D.N.K. and M.K.G.) are
thankful to the Bulgarian National Science Fund for partial
support under the Contracts $\Phi$-1416 and $\Phi$-1501. This work
was partly supported by the Agreement (2004 BG2004) between the
CSIC (Spain) and the Bulgarian Academy of Sciences, by the
Agreement between JINR (Dubna) and INRNE (Sofia) and by funds
provided by MEC (Spain) under Contracts BFM 2002-03562 and
2003-04147-C02-01.


\begin{thebibliography}{99}

% 1.
\bibitem{Tan85a} I. Tanihata, H. Hamagaki, O. Hashimoto, S. Nagamiya,
Y. Shida, N. Yoshikawa, O. Yamakawa, K. Sugimoto, T. Kobayashi, D.
E. Greiner, N. Takahashi, and Y. Nojiri, Phys. Lett.
\textbf{B160}, 380 (1985).

% 2.
\bibitem{Tan85b} I. Tanihata, H. Hamagaki, O. Hashimoto, Y. Shida, N.
Yoshikawa, K. Sugimoto, O. Yamakawa, and T. Kobayashi, Phys. Rev.
Lett. \textbf{55}, 2676 (1985).

% 3.
\bibitem{Tan85c} I. Tanihata, Prog. Part. Nucl. Phys. \textbf{35}, 505
(1985).

% 4.
\bibitem{Tan88a} I. Tanihata, Nucl. Phys. \textbf{A488},
113c (1988).

% 5.
\bibitem{Tan88b} I. Tanihata, T. Kobayashi, O. Yamakawa, S. Shimoura,
K. Ekuni, K. Sugimoto, N. Takahashi, T. Shimoda, and H. Sato,
Phys. Lett. \textbf{B206}, 592 (1988).

% 6.
\bibitem{Tan92} I. Tanihata, D. Hirata, T. Kobayashi, S. Shimoura, K.
Sugimoto, and H. Toki, Phys. Lett. \textbf{B289}, 261 (1992).

% 7.
\bibitem{Han95} P. G. Hansen, A. S. Jensen, and B. Jonson, Ann. Rev.
Nucl. Sci. \textbf{45}, 591 (1995).

% 8.
\bibitem{Dob94} J. Dobaczewski, I. Hamamoto, W. Nazarewicz, and J. A.
Sheikh, Phys. Rev. Lett. \textbf{72}, 981 (1994).

% 9.
\bibitem{Cas00} R. F. Casten and B. M. Sherill, Prog. Part. Nucl.
Phys. \textbf{45}, S171 (2000), see also the special issue of
Nucl. Phys. \textbf{A693}, Nos. 1-2 (2001).

\bibitem{Tani95} I. Tanihata, Prog. Part. Nucl.
Phys. \textbf{35}, 505 (1995).

% 10.
\bibitem{Oga99a} Yu. T. Oganessian, V. I. Zagrebaev, and J. S. Vaagen,
Phys. Rev. Lett. \textbf{82}, 4996 (1999).

% 11.
\bibitem{Oga99b} Yu. T. Oganessian, V. I. Zagrebaev, and J. S. Vaagen,
Phys. Rev. C \textbf{60}, 044605 (1999).

% 12.
\bibitem{Cor97} M. D. Cortina-Gil, P. Roussel-Chomaz, N. Alamanas, J.
Barette, W. Mittig, F. Auger, Y. Blumenfeld, J. M. Casandjian, M.
Chartier, V. Fekou-Youmbi, B. Fernandez, N. Frascaria, A.
Gillibert, H. Laurent, A. Lepine-Szily, N. A. Orr, V. Pascalon, J.
A. Scarpaci, J. L. Sida, and T. Saomijarvi, Nucl. Phys.
\textbf{A616}, 215c (1997).

% 13.
\bibitem{Lag01} A. Lagoyannis, F. Auger, A. Mussumara, N. Alamanos, E.
C. Polacco, A. Pakou, Y. Blumenfeld, F. Braga, M. L. Commara, A.
Drouart, G. Fioni, A. Gillebert, E. Khan, V. Lapoux, W. Mittig, S.
Ottini-Hustache, D. Pierroutsakou, M. Romoli, P. Roussel-Chomaz,
M. Sandoli, D. Santonocito, J. A. Scarpaci, J. L. Sida, T.
Saomijarvi, S. Karataglidis, and K. Amos, Phys. Lett.
\textbf{B518}, 27 (2001).

% 14.
\bibitem{Cor96} M. D. Cortina-Gil, P. Roussel-Chomaz, N. Alamanas, J.
Barette, W. Mittig, F. Auger, Y. Blumenfeld, J. M. Casandjian, M.
Chartier, V. Fekou-Youmbi, B. Fernandez, N. Frascaria, A. Gillibert,
H. Laurent, A. Lepine-Szily, N. A. Orr, V. Pascalon, J. A. Scarpaci,
J. L. Sida, and T. Saomijarvi, Phys. Lett. \textbf{B371}, 14 (1996).

% 15.
\bibitem{Kor97a} A. A. Korsheninnikov, E. A. Kuzmin, E. Yu. Nikolskii,
C. A. Bertulani, O. V. Bochkarev, S. Fukuda, T. Kobayashi, S. Momota,
B. G. Novatskii, A. A. Ogloblin, A. Ozawa, V. Pribora, I. Tanihata,
and K. Yoshida, Nucl. Phys. \textbf{A616}, 189c (1997).

% 16.
\bibitem{Kor97b} A. A. Korsheninnikov, E. Yu. Nikolskii, C. A.
Bertulani, S. Fukuda, T. Kobayashi, E. A. Kuzmin, S. Momota, B. G.
Novatskii, A. A. Ogloblin, A. Ozawa, V. Pribora, I. Tanihata, and K.
Yoshida, Nucl. Phys. \textbf{A617}, 45 (1997).

% 17.
\bibitem{Ter01} G. M. Ter-Akopian \textit{et al.}, in
\textit{Fundamental issues in elementary}, Proc. Sym. Honor and
Memory of Michael Danos, Bad Honnef, Germany, 2000, edited by W.
Greiner (EP Systema, Debrecen, 2001), p. 371.

% 18.
\bibitem{Chu95} L. L. Chulkov, C. A. Bertulani, and A. A.
Korsheninnikov, Nucl. Phys. \textbf{A587}, 291 (1995).

% 19.
\bibitem{Kor93} A. A. Korsheninnikov, K. Yoshida, D. V. Alexandrov, N.
Aoi, Y. Doki, N. Inabe, M. Fujimaki, T. Kobayashi, H. Kumagai,
C.-B. Moon, E. Yu. Nikolskii, M. M. Obuti, A. A. Ogloblin, A.
Ozawa, S. Shimoura, Y. Watanabe, and M. Yanokura, Phys. Lett.
\textbf{B316}, 38 (1993).

% 20.
\bibitem{Alk02} G. D. Alkhazov, A. V. Dobrovolsky, P. Edelhof, H.
Geissel, H. Irnich, A. V. Khanzadeev, G. A. Korolev, A. A.
Lobodenko, G. M\"{u}nzenberg, M. Mutterer, S. R. Neumaier, W.
Schwab, D. M. Selivestrov, T. Suzuki, and A. A. Vorobyov, Nucl.
Phys. \textbf{A712}, 269 (2002).

% 21.
\bibitem{Ege01} P. Egelhof, Prog. Part. Nucl. Phys. \textbf{46}, 307
(2001).

% 22.
\bibitem{Ege02} P. Egelhof \textit{et al.}, Eur. Phys. J.
\textbf{A15}, 27 (2002).

% 23.
\bibitem{Neu02} S. R. Neumeier \textit{et al.}, Nucl. Phys.
\textbf{A712}, 247 (2002).

% 24.
\bibitem{Ege03} P. Egelhof, O. Kisselev, G. M\"{u}nzenberg, S. R. Neumeier,
and H. Weick, Physica Scripta \textbf{T104}, 151 (2003) (and
references therein).

% 25.
\bibitem{Cre95} R. Crespo, J. A. Tostevin, and R. C. Johnson, Phys.
Rev. C \textbf{51}, 3283 (1995).

% 26.
\bibitem{Zhu93} M. V. Zhukov, B. V. Danilin, D. V. Fedorov, J. M.
Bang, I. J. Thompson, and J. S. Vaagen, Phys. Rep. \textbf{231}, 151
(1993).

% 27.
\bibitem{Zhu94} M. V. Zhukov, A. A. Korsheninnikov, and M. H.
Smedberg, Phys. Rev. C \textbf{50}, R1 (1994).

% 28.
\bibitem{Avr00} M. Avrigeanu, G. S. Anagnostatos, A. N. Antonov, and
J. Giapitzakis, Phys. Rev. C \textbf{62}, 017001 (2000).

% 29.
\bibitem{Avr02} M. Avrigeanu, G. S. Anagnostatos, A. N. Antonov, and
V. Avrigeanu, Int. J. Mod. Phys. \textbf{E11}, 249 (2002).

% 30.
\bibitem{Dor98} P. J. Dortmans, K. Amos, S. Karataglidis, and J.
Rainal, Phys. Rev. C \textbf{58}, 2249 (1998).

% 31.
\bibitem{Kar97} S. Karataglidis, B. A. Brown, K. Amos, and P. J.
Dortmans, Phys. Rev. C \textbf{55}, 2826 (1997).

% 32.
\bibitem{NuP00} Radioactive Nuclear Beam Facilities, NuPECC Report
(Europe, 2000).

% 33.
\bibitem{Shr01} Electromagnetic Probes and the Structure of Hadrons
and Nuclei, Prog. Part. Nucl. Phys. \textbf{44} (2000); G.
Shrieder \textit{et al.}, Proposal for GSI RI Beam Factory (2001).

% 34.
\bibitem{Sud01} T. Suda, in \textit{Challenges of Nuclear Structure},
Proceedings of 7th International Spring Seminar on Nuclear Physics,
edited by Aldo Covello (World Scientific Publishing Co., Singapore,
2002), p. 13; Proposal for RIKEN RI Beam Factory (2001).

% 35.
\bibitem{Kat03} T. Katayama, T. Suda, and I. Tanihata, Physica
Scripta, \textbf{T104}, 129 (2003).

% 36.
\bibitem{Gar99} E. Garrido and E. Moya de Guerra, Nucl. Phys.
\textbf{A650}, 387 (1999).

% 37.
\bibitem{Gar00} E. Garrido and E. Moya de Guerra, Phys. Lett.
\textbf{B488}, 68 (2000).

% 38.
\bibitem{Moy02} E. Moya de Guerra, E. Garrido, and P. Sarriguren,
in \textit{Challenges of Nuclear Structure}, Proceedings of 7th
International Spring Seminar on Nuclear Physics, edited by Aldo
Covello (World Scientific Publishing Co., Singapore, 2002), p. 63.

% 39.
\bibitem{Ant04} A. N. Antonov, M. K. Gaidarov, D. N. Kadrev, P. E.
Hodgson, and E. Moya de Guerra, Int. J. Mod. Phys. \textbf{E13}, 759
(2004).

% 40.
\bibitem{Kar00} S. Karataglidis, P. J. Dortmans, K. Amos, and C.
Bennhold, Phys. Rev. C \textbf{61}, 024319 (2000).

% 41.
\bibitem{Suz99} T. Suzuki, R. Kanungo, O. Bochkarev, L. Chulkov, D.
Cortina, M. Fukuda, H. Geissel, M. Hellstr\"{o}m, M. Ivanov, R.
Janik, K. Kimura, T. Kobayashi, A. A. Korsheninnikov, G.
M\"{u}nzenberg, F. Nickel, A. A. Ogloblin, A. Ozawa, M.
Pf\"{u}tzner, V. Pribora, H. Simon, B. Sit\`{a}r, P. Strmen, K.
Sumiyoshi, K. Summerer, I. Tanihata, M. Winkler, and K. Yoshida,
Nucl. Phys. \textbf{A658}, 313 (1999).

% 42.
\bibitem{Wang04} Z. Wang and Z. Ren, Phys. Rev. C \textbf{70}, 034303
(2004); {\it ibid.} \textbf{71}, 054323 (2005).

% 43.
\bibitem{Richter03} W. A. Richter and B. A. Brown, Phys. Rev. C \textbf{67},
034317 (2003).

% 44.
\bibitem{AlKh96} J. S. Al-Khalili, J. A. Tostevin, and I. J. Thompson,
Phys. Rev. \textbf{C54}, 1843 (1996).

% 45.
\bibitem{Tost97} J. A. Tostevin and J. S. Al-Khalili, Nucl. Phys.
\textbf{A616}, 418c (1997).

% 46.
\bibitem{SarXX} P. Sarriguren \textit{et al.}, in preparation.

% 47.
\bibitem{Sar99} P. Sarriguren, E. Moya de Guerra, and A. Escuderos,
Nucl. Phys. \textbf{A658}, 13 (1999).

% 48.
\bibitem{Yen54} D. R. Yennie, D. G. Ravenhall, and R. N. Wilson, Phys.
Rev. \textbf{95}, 500 (1954) (and references therein).

% 49.
\bibitem{Luk02} V. K. Lukyanov, E. V. Zemlyanaya, D. N. Kadrev, A.
N. Antonov, K. Spasova, G. S. Anagnostatos, and J. Giapitzakis,
Particles and Nuclei, Letters, No. 2[111], 5 (2002); Bull. Rus.
Acad. Sci., Phys. \textbf{67}, 790 (2003).

% 50.
\bibitem{Nis85} M. Nishimura, E. Moya de Guerra, and D. W. L. Sprung,
Nucl. Phys. \textbf{A435}, 523 (1985); J. M. Udias, M.Sc. Thesis,
Universidad Autonoma de Madrid (unpublished) (1987).

% 51.
\bibitem{For66} T. de Forest, Jr. and J. D. Walecka, Adv. Phys.
\textbf{15}, 1 (1966).

% 52.
\bibitem{Alk91} G. D. Alkhazov, V. V. Anisovich, and P. E. Volkovyckii,
\textit{Diffractional Interaction of Hadrons with Nuclei at High
Energies} (Nauka, Leningrad, 1991), p. 94.

% 53.
\bibitem{Bur77} V. V. Burov and V. K. Lukyanov, Preprint JINR,
R4-11098, 1977, Dubna; V. V. Burov, D. N. Kadrev, V. K. Lukyanov,
and Yu. S. Pol', Phys. At. Nuclei \textbf{61}, No.4, 525 (1998).

\bibitem{Friedrich03} J. Friedrich and Th. Walcher, Eur. Phys. J.
A \textbf{17}, 607 (2003).

% 54.
\bibitem{Sim80} G. G. Simon, Ch. Schmitt, F. Borkowski, and V. H. Walther,
Nucl. Phys. \textbf{A333}, 381 (1980).

\bibitem{Galster71} S. Galster, H. Klein, J. Moritz, K. H. Schmidt,
and D. WegenerJ. Bleckwenn, Nucl. Phys. \textbf{B32}, 221 (1971).

% 55.
\bibitem{Chan76} H. Chandra and G. Sauer, Phys. Rev. C \textbf{13},
245 (1976).

% 57.
\bibitem{Vri87} H. De Vries, C. W. De Jager, and C. De Vries, Atomic
Data and Nuclear Data Tables \textbf{36}, 495 (1987).

% 58.
\bibitem{Sick82} I. Sick, Phys. Lett. \textbf{B116}, 212 (1982).

% 59.
\bibitem{Pat03} J. D. Patterson and R. J. Peterson, Nucl. Phys.
\textbf{A717}, 235 (2003).

% 60.
\bibitem{Fri94} E. Friedman, A. Gal, and J. Mares, Nucl. Phys.
\textbf{A579}, 518 (1994).

% 61.
\bibitem{Vautherin} D. Vautherin, Phys. Rev. C {\bf 7}, 296 (1973).

\bibitem{GSI2005} {\it Technical Proposal for the Design,
Construction, Commisioning and Operation of the ELISe Setup},
Spokesperson: H. M. Simon (2005).

\bibitem{McCarthy77} J. S. McCarthy, I. Sick, and R. R. Whitney, Phys. Rev.
C {\bf 15}, 1396 (1977).

\bibitem{Otter85} C. R. Ottermann \textit{et al.}, Nucl. Phys. \textbf{A436}, 688
(1985).

\bibitem{Suelzle67} L. R. Suelzle, M. R. Yearian, and Hall Crannell, Phys. Rev. C
{\bf 162}, 992 (1967).

\bibitem{Li71} G. C. Li, I. Sick,  R. R. Whitney, and M. R. Yearian, Nucl.
Phys. \textbf{A162}, 583 (1971).

\bibitem{Sick75} I. Sick, J. B. Bellicard, M. Bernheim, B. Frois, M. Huet, Ph.
Leconte, J. Mougey, Phan Xuan-Ho, D. Royer, and S. Turck, Phys.
Rev. Lett. {\bf 35}, 910 (1975).

% 62.
\bibitem{Lit71} A. S. Litvinenko \textit{et al.}, Yad. Fiz.
\textbf{14}, 40 (1971) (Sov. J. Nucl. Phys.\textbf{14}, 23 (1972))

% 63.
\bibitem{Lit72} A. S. Litvinenko \textit{et al.}, Nucl. Phys.
\textbf{A182}, 265 (1972).

\bibitem{Curtis69} T. H. Curtis, R. A. Eisenstein, D. W. Madsen, and C. K.
Bockelman, Phys. Rev. {\bf 184}, 1162 (1969).

\bibitem{Light76} J. W. Lightbody, Jr., S. Penner, S. P. Fivozinsky, P. L.
Hallowell, and H. Crannell, Phys. Rev. C {\bf 14}, 952 (1976).

\bibitem{Cavedon82} J. Cavedon, J. B. Bellicard, B. Frois, D. Goutte, M. Huet, P.
Leconte, X.-H. Phan, S. K. Platchkov, and I. Sick, Phys. Lett.
{\bf B118}, 311 (1982).

\bibitem{Ube71} H. Uberall, \textit{Electron Scattering from
Complex Nuclei} (Academic Press, New York, 1971).

% 64
\bibitem{Keim95} M. Keim \textit{et al.}, Nucl. Phys.
{\bf A586}, 219 (1995).

% 65.
\bibitem{Abbott2000} D. Abbott \textit{et al.}, Eur. Phys. J.
\textbf{A7}, 421 (2000).

% 66.
\bibitem{Ozawa2001} A. Ozawa, T. Suzuki, and I. Tanihata, Nucl. Phys.
\textbf{A693}, 32 (2001).

% 67.
\bibitem{Swia05} W. J. Swiatecki, A. Trzcinska, and J. Jastrzebski,
Phys. Rev. C \textbf{71}, 047301 (2005).

\end{thebibliography}
\end{document}